\documentclass[aps,prd,preprint,eqsecnum,nofootinbib,groupedaddress,floatfix,12pt]{revtex4}

\usepackage{graphicx,amssymb,amsbsy,amsfonts,amssymb,amsmath}
\usepackage{hyperref}
\usepackage{multirow}
\usepackage{stackrel}
\usepackage{tikz}
\usetikzlibrary{calc}

\newcommand{\be}{\begin{equation}}
\newcommand{\ee}{\end{equation}}

\newcommand{\bea}{\begin{eqnarray}}
\newcommand{\eea}{\end{eqnarray}}

\newcommand{\eqnDiag}[1]{ \vcenter{\hbox{#1}} }

\def\sect#1{section~{\ref{#1}}}

\def\eqn#1{eq.~(\ref{#1})}
\def\eqns#1#2{eqs.~(\ref{#1}) and (\ref{#2})}
\def\fig#1{fig.~{\ref{#1}}}

\def\spa#1.#2{\left\langle#1\,#2\right\rangle}
\def\spb#1.#2{\left[#1\,#2\right]}
\def\spash#1.#2{\spa{\smash{#1}}.{\smash{#2}}}
\def\spbsh#1.#2{\spb{\smash{#1}}.{\smash{#2}}}
\def\sand#1.#2.#3{%
\left\langle\smash{#1}{\vphantom1}^{-}\right|{#2}%
\left|\smash{#3}{\vphantom1}^{-}\right\rangle}
\def\sandpp#1.#2.#3{%
\left\langle\smash{#1}{\vphantom1}^{+}\right|{#2}%
\left|\smash{#3}{\vphantom1}^{+}\right\rangle}
\def\sandpm#1.#2.#3{%
\left\langle\smash{#1}{\vphantom1}^{+}\right|{#2}%
\left|\smash{#3}{\vphantom1}^{-}\right\rangle}
\def\sandmp#1.#2.#3{%
\left\langle\smash{#1}{\vphantom1}^{-}\right|{#2}%
\left|\smash{#3}{\vphantom1}^{+}\right\rangle}

\def\ibp{IBP}

\def\BlackHat{{\sc BlackHat}}

\newbox\charbox
\newbox\slabox
\def\s#1{{      
        \setbox\charbox=\hbox{$#1$}
        \setbox\slabox=\hbox{$/$}
        \dimen\charbox=\ht\slabox
        \advance\dimen\charbox by -\dp\slabox
        \advance\dimen\charbox by -\ht\charbox
        \advance\dimen\charbox by \dp\charbox
        \divide\dimen\charbox by 2
        \raise-\dimen\charbox\hbox to \wd\charbox{\hss/\hss}
        \llap{$#1$} }}

\usepackage{young}

\begin{document}

\hbox{\rm\small
FR-PHENO-2017-002$\null\hskip 8.5cm \null$
\break}

\title{Subleading Poles in the Numerical Unitarity Method at Two Loops}

\author{
    S.~Abreu, F.~Febres Cordero, H.~Ita, M.~Jaquier and B.~Page
\\
$\null$
\\
Physikalisches Institut, Albert-Ludwigs-Universit\"at Freiburg\\
       D--79104 Freiburg, Germany \\
}

\begin{abstract}
We describe the unitarity approach for the numerical computation of two-loop
integral coefficients of scattering amplitudes. It is well known that the leading
propagator singularities of an amplitude's integrand are related to products
of tree amplitudes.  At two loops, Feynman diagrams with doubled
propagators appear naturally, which lead to subleading pole contributions.
In general, it is not known how these contributions can be directly expressed in
terms of a product of on-shell tree amplitudes. 
We present a universal algorithm to extract these subleading pole terms by releasing
some of the on-shell conditions. 
We demonstrate the new approach by numerically computing  two-loop four-gluon
integral coefficients.
\end{abstract}

\maketitle

\numberwithin{equation}{section}

\section{Introduction} 
\label{sec:Introduction}

The unitarity method~\cite{Unitarity} constructs scattering amplitudes
from their unitarity and analytic structure. It is convenient to work
at the integrand level where factorization properties tie  the
leading coefficients of the propagator poles to products of tree amplitudes.
In the last decade, numerical approaches~\cite{OPP,NumUnitarity,BlackHat,GKM}
have been developed that construct one-loop amplitudes from
their propagator poles, which are given by tree amplitudes. 
Beyond one loop, Feynman rules naturally yield contributions with
higher-order propagator powers. In order to determine the rational integrand,
one thus has to obtain leading and subleading coefficients on such propagator
poles.  Of these, only the leading ones are directly related to a
product of tree amplitudes and the subleading terms have to be obtained
differently.
For analytic computations this obstruction has been discussed in
\cite{DPropAnalytic} where the residue extraction is adjusted to pick up
subleading-pole contributions.
Alternatively, in analytic computations of two-loop QCD
amplitudes~\cite{fourgluons}, the subleading-pole contributions can be tracked
explicitly and evaluated or, in the case of particular helicity amplitudes,
dealt with by choosing particular representations of the
integrand~\cite{UnitarityAnalytics2loop}. 

In this article we propose a numerical algorithm to extract subleading-pole
contributions without resorting to analytic manipulations.  The central idea
is to `cut less', i.e. to obtain subleading-pole contributions from their
contribution to cuts which keep the respective propagators off-shell.  
We find that the algorithm works effectively when applied to a numerical
calculation of two-loop four-gluon amplitudes. We validate our approach
by comparing it to the known analytic results~\cite{fourgluons},
and by carrying out a number of non-trivial consistency checks.

The rest of the paper is organized as follows. In Section~\ref{sec:setup} we
describe the organization of a calculation in the numerical unitarity
method, discuss the appearance of
subleading-pole terms starting at two-loops, and present our algorithm for
extracting those terms.
Section~\ref{sec:applications} contains applications in the context of
one- and two-loop four-gluon amplitudes as well as a list of checks performed.
Finally, in Section~\ref{sec:outlook} we present our conclusions and outlook.

\section{Setup for two-loop numerical unitarity}\label{sec:setup}

In this section, we review the main aspects of the numerical unitarity method at
two loops. We introduce our notation, explain the appearance of
subleading-pole terms at two loops and present our algorithm to extract them. Although we
focus on two-loop calculations, our result can be easily generalized for
computing general subleading-pole contributions in multi-loop amplitudes.

\subsection{Diagrammatic decomposition}

\begin{figure}[]
\includegraphics[scale=.7]{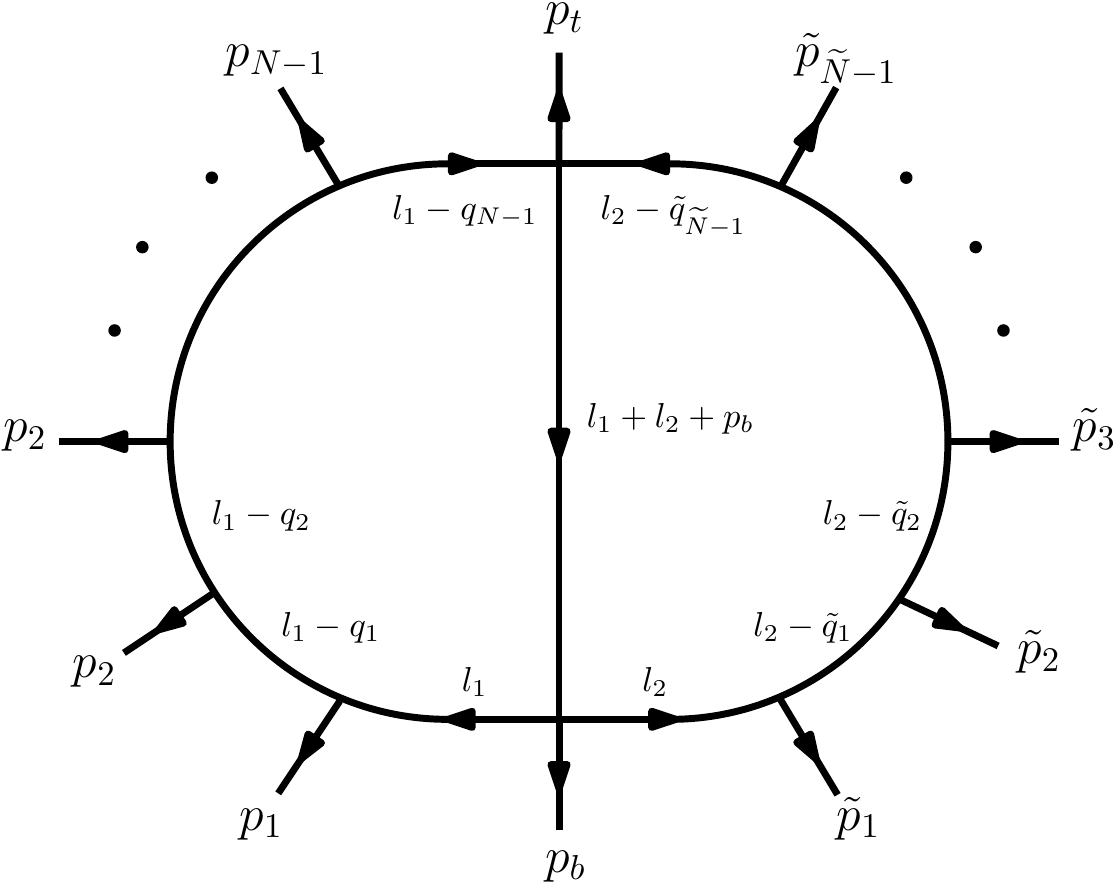}
\caption{A generic diagram depicting the propagator structure that appears in a
two-loop planar amplitude. The momenta $q_i$
and $\tilde q_i$ are determined by momentum conservation.}
\label{fig:2loopdiag} 
\end{figure}

We will organize amplitudes in terms of diagrams, which can be constructed
from the usual Feynman diagram decomposition in the following way: we strip
Feynman diagrams of particle information and pinch all propagators through which loop
momentum does not flow.  We will denote the set of all diagrams constructed in
this way by $\Delta$. In the remainder of this paper, a diagram $\Gamma$
denotes an element of this set, $\Gamma\in \Delta$.  Each $\Gamma$ defines a
set of propagator indices $P_\Gamma$, and we call the set of inverse
propagators $\{\rho_k\}$ with $k\in P_\Gamma$ a {\it propagator structure}. As
we allow for elements in $P_\Gamma$ to be repeated, technically we should call
$P_\Gamma$ a  multiset, but we avoid this terminology.  In \fig{fig:2loopdiag}
we show a generic diagram $\Gamma$ for a planar two-loop amplitude.  
Finally, we associate the notion of {\it hierarchy} to $\Delta$. 
If $\Gamma_1$ and $\Gamma_2$ are two elements of $\Delta$, such that $\Gamma_2$
is obtained by pinching some of the edges of $\Gamma_1$, then they are members
of the same hierarchy.  $\Gamma_1$ is called an ancestor of $\Gamma_2$, and we
write $\Gamma_1>\Gamma_2$ to denote ancestry.
All of the descedant's propagators are contained in any of its ancestors,
i.e.~$P_{\Gamma_2}\subset P_{\Gamma_1}$.
Different
hierarchies in $\Delta$ are named according to their element with the fewest
edges.

\subsection{Master integrals and integrands}

The scattering amplitude~${\cal A}$ is decomposed in the general form
\begin{equation}
    {\cal A}=\sum_{\Gamma\in \Delta}\sum_{i\,\in\,M_\Gamma} c_{\Gamma,i}\, I_{\Gamma,i}\,,
\label{eq:A}
\end{equation}
in terms of a set of {\it master integrals} $I_{\Gamma,i}$ and
coefficient functions $c_{\Gamma,i}$. 
We organize the sum over master integrals according to their propagator
structure labeled by $\Gamma$. By $M_\Gamma$ we denote the set of indices $i$
which label the different master integrals that share the same propagator structure.
$M_\Gamma$ can be empty if no master integral is associated with diagram $\Gamma$.
We work in dimensional regularization, so that the integrals and their
coefficients depend on the space-time dimension $D$. In addition, the integral
coefficients depend on the dimensionality $D_s$ of the spin space for the loop
particles~\cite{FDH}. For simplicity we neither display the dimensional dependence
nor the natural dependence on kinematic variables. 
Furthermore, for the present discussion it is sufficient to consider fixed
values of $D$ and $D_s$ for which the amplitude is finite. 
Also, we always consider planar amplitudes, as this is sufficient for our purposes in
this paper (notice that for two-loop diagrams higher propagator powers appear
only in planar amplitudes).

In a numerical approach, it is helpful to analyze \eqn{eq:A} prior to
integrating over loop momenta, i.e., to analyze the integrand ${\cal
A}(\ell_l)$. 
The symbol $\ell_l$ represents the momenta of the two loops
and will be used to denote quantities defined at the integrand level.  
The integrand is decomposed into terms that
contribute to the sum in \eqn{eq:A}, which we call {\it master integrands},
and independent {\it surface integrands} which integrate to
zero~\cite{NumUnitarity2},
\begin{equation}\label{eq:AL} {\cal A}(\ell_l)=
    \sum_{\Gamma\in \Delta} \frac{1}{ \prod_{k\in P_\Gamma} \rho_k }\,\,
    \sum_{ i\,\in\,M_\Gamma\cup S_\Gamma} c_{\Gamma,i}\,m_{\Gamma,i}(\ell_l)\,, 
\end{equation}
where $M_\Gamma$ and $S_\Gamma$ denote the set of master integrands and surface
terms associated to diagram $\Gamma$, respectively. 
The numerator terms $m_{\Gamma,i}(\ell_l)$
integrate either to master integrals or to zero, 
\begin{equation} 
\int  \frac{d^D\ell_1d^D\ell_2}{(2\pi)^{2D}}\,
\frac{m_{\Gamma,i}(\ell_l)}{\prod_{k\in P_\Gamma} \rho_k} = \left\{
    \begin{array}{cc} I_{\Gamma,i} &  \mbox{for}\quad i\in
M_\Gamma\,, \\ 0&  \mbox{for}\quad i\in S_\Gamma\,. \end{array}\right.
\end{equation}
Similar notation has been used for example in~\cite{Badger:2016ozq}. 
For future reference, we define the integrand numerator
${N}(\Gamma,\ell_l)$ associated with the propagator
structure~$\Gamma$ by
\begin{align}\label{eq:N}
    {N}^{}(\Gamma,\ell_l)&= \sum_{ i\,\in\,{M_\Gamma}\cup{S_\Gamma}} c_{\Gamma,i}
    \,m_{\Gamma,i}(\ell_l)\,.
\end{align} 
The construction of the integrand representation in \eqn{eq:AL} has been given
in \cite{NumUnitarity2}, using appropriate integration-by-parts (\ibp)
identities~\cite{IBPGKK}.  
The \ibp{} relations have to be chosen sufficiently general in order to include
the propagator structures already present in the integrand ${\cal A}(\ell_l)$
of the amplitude, e.g. given by Feynman rules.  In particular, given that two-loop amplitudes
contain diagrams with doubled propagators, we have to consider such propagator
structures as well when constructing the sets of master integrands and surface terms.

\subsection{Integrand coefficients and factorization}
\label{sec:genUni}

The coefficient functions $c_{\Gamma,i}$, can be obtained by
solving the linear system of equations (\ref{eq:AL}) for generic values of the
loop momentum. In the generalized unitarity approach, the system of equations
is analyzed diagram by diagram.  For each diagram $\Gamma$, we consider the specific values
of the loop momenta $\ell_l^\Gamma$ where internal particles go on-shell,
\begin{eqnarray} \ell_l^\Gamma:\quad \ell_l\quad\mbox{with}\quad\rho_k=0 \quad \mbox{for all}
\quad k\in P_\Gamma\,. \end{eqnarray}
In the limit $\ell_l\rightarrow\ell_l^\Gamma$, both sides of \eqn{eq:AL} diverge and the coefficients of the poles can be compared,
yielding a refined system of equations. Importantly, unitarity and
factorization properties of field theory amplitudes imply that the leading coefficients of
the poles in ${\cal A}(\ell_l)$ are given by products of tree amplitudes.

\begin{figure}[]
    \begin{tikzpicture}[scale=5]
    \node at (0,2){\includegraphics[scale=0.35]{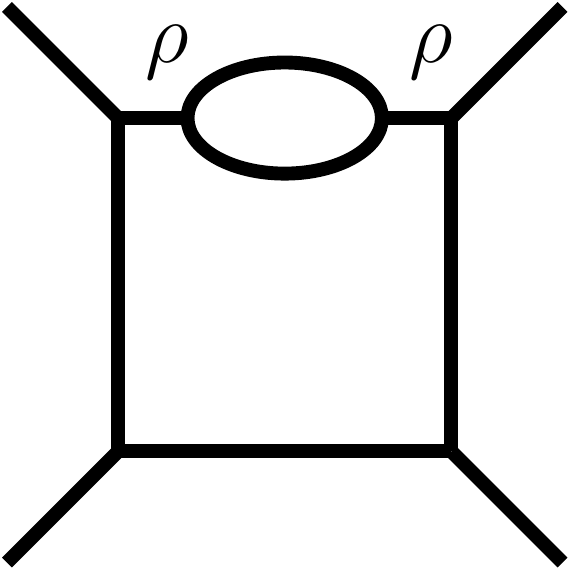}};
    \node at (0,1.7){(a)};
    \node at (1,2){\includegraphics[scale=0.35]{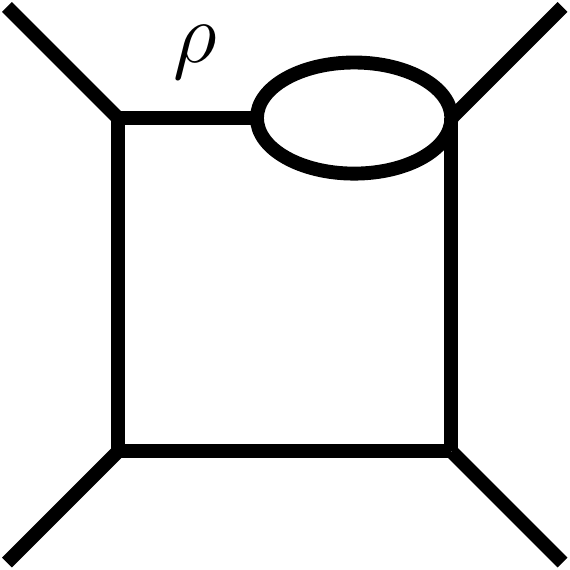}};
    \node at (1,1.7){(b)};
    \end{tikzpicture}
    \caption{ Two diagrams with the
    same set of propagators. Propagator $1/\rho$ appears twice in diagram~(a) but only once
    in diagram (b).}
    \label{fig:DProptopologies}
\end{figure}

\begin{figure}[ht]
\begin{tikzpicture}[scale=2.2]
    \draw[dashed] (5.3,2.7) -- (5.3,0.22);
    \node at (1,3.4){\includegraphics[scale=0.22]{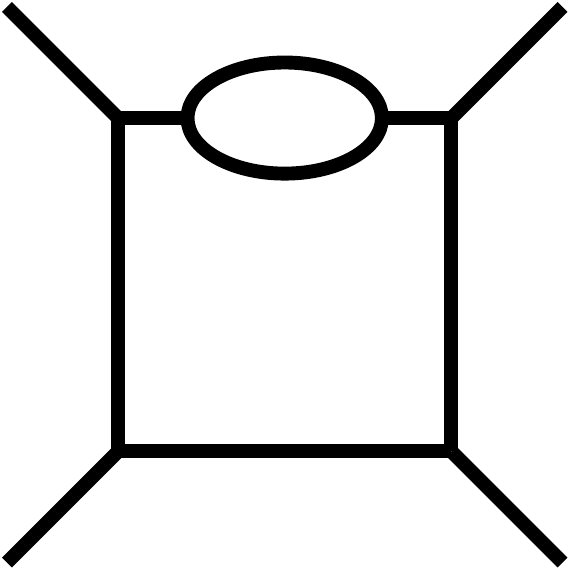}};
    \node at
    (2.55,3.4){\includegraphics[scale=0.22]{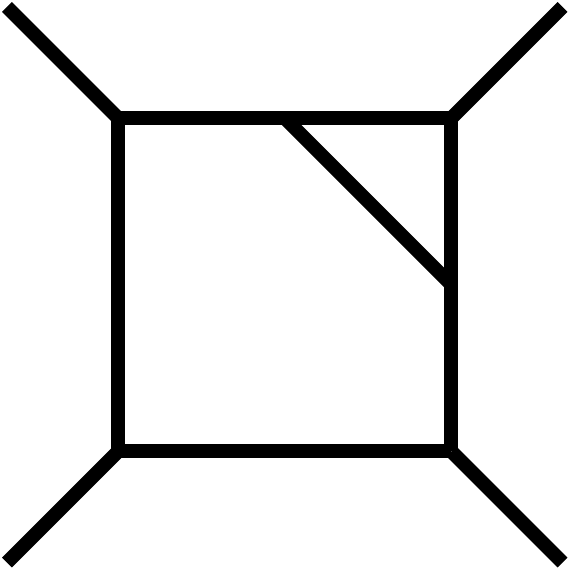}}; \node
    at (4.1,3.4){\includegraphics[scale=0.22]{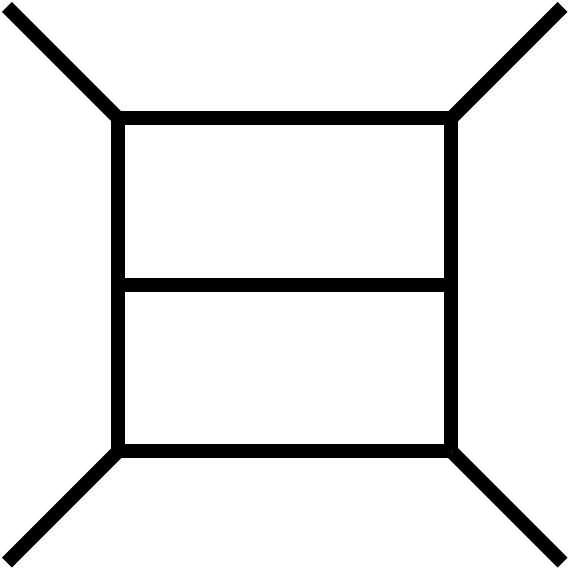}};
    \node at
    (0.55,2.4){\includegraphics[scale=0.22]{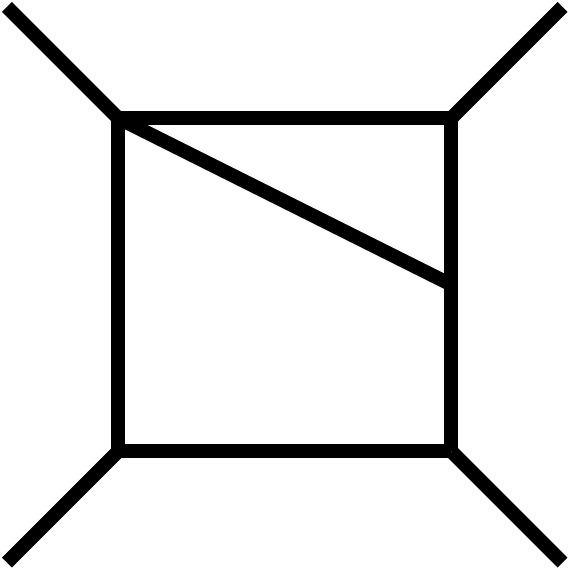}}; \node
    at (1.55,2.4){\includegraphics[scale=0.22]{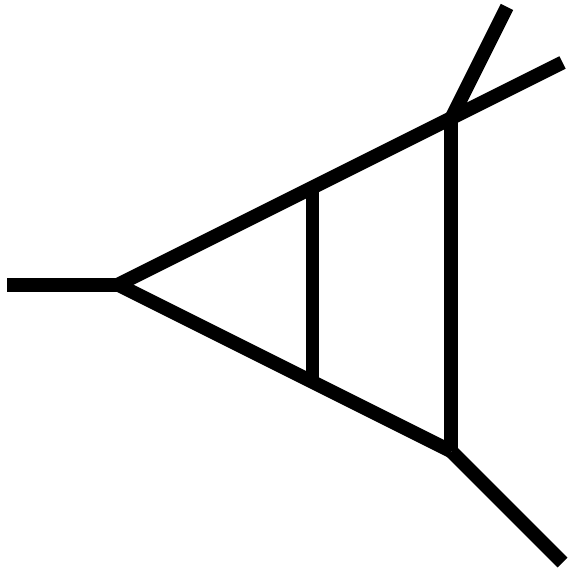}};
    \node at
    (2.55,2.4){\includegraphics[scale=0.22]{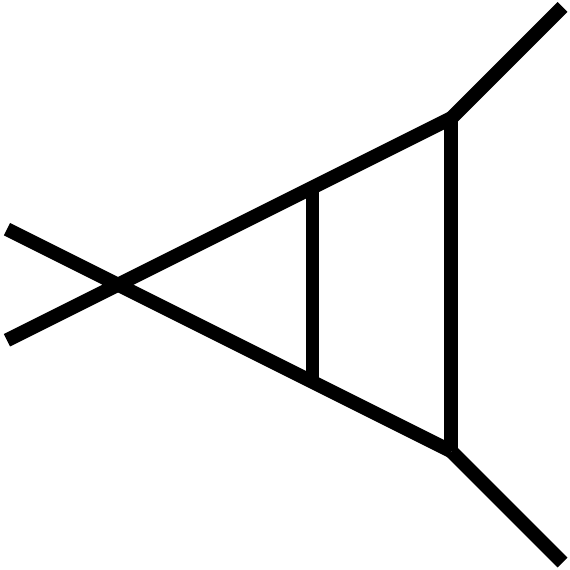}};
    \node at
    (3.55,2.4){\includegraphics[scale=0.22]{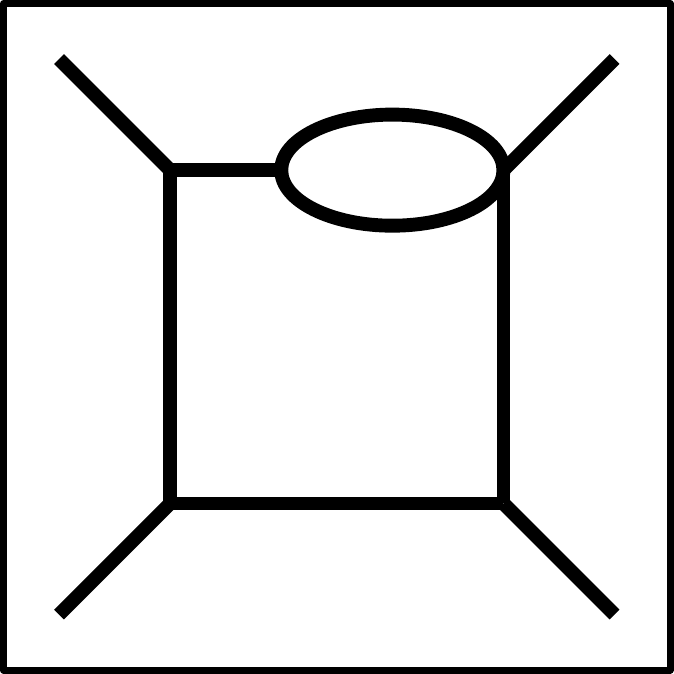}};
    \node at
    (4.55,2.4){\includegraphics[scale=0.22]{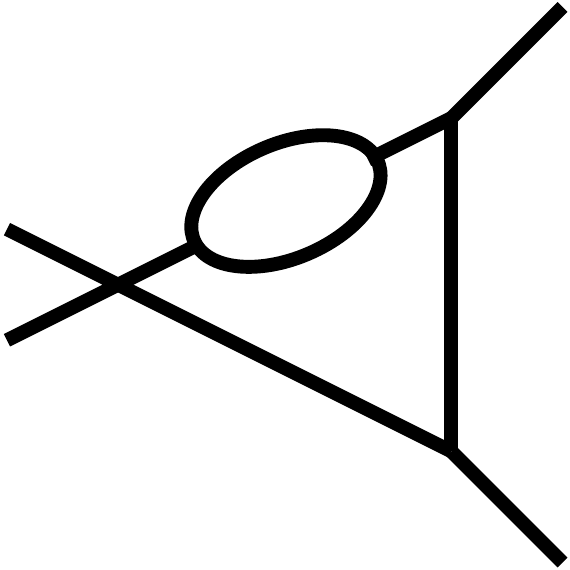}};
    \node at
    (5.8,2.4){\includegraphics[scale=0.22]{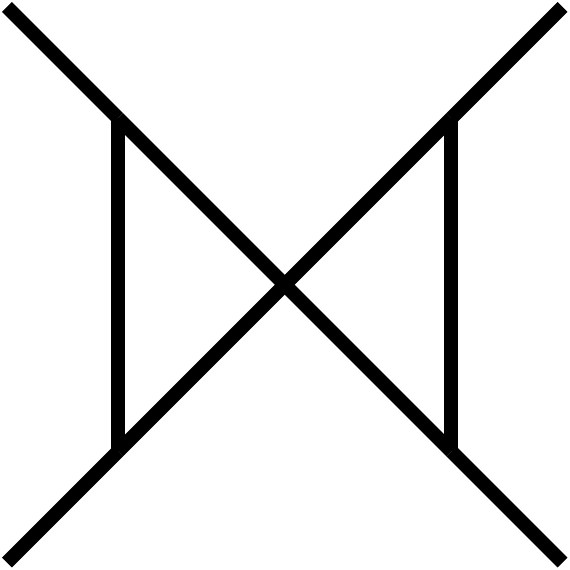}};
    \node at
    (6.5,2.4){\includegraphics[scale=0.22]{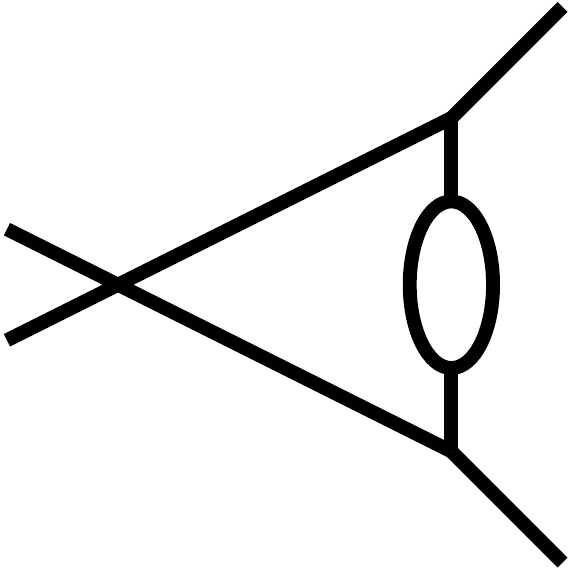}};
    \node at
    (0,1.4){\includegraphics[scale=0.22]{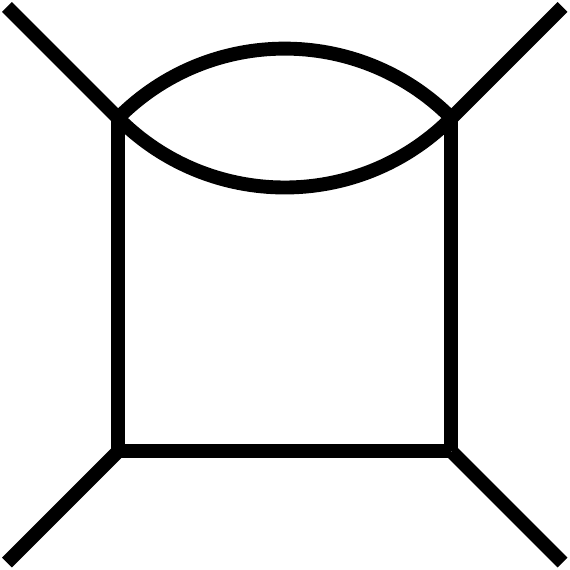}}; \node
    at
    (.8,1.4){\includegraphics[scale=0.22]{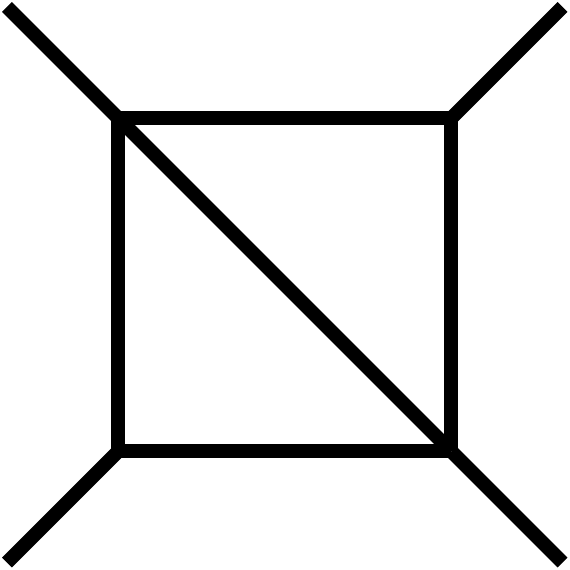}};
    \node at
    (1.6,1.4){\includegraphics[scale=0.22]{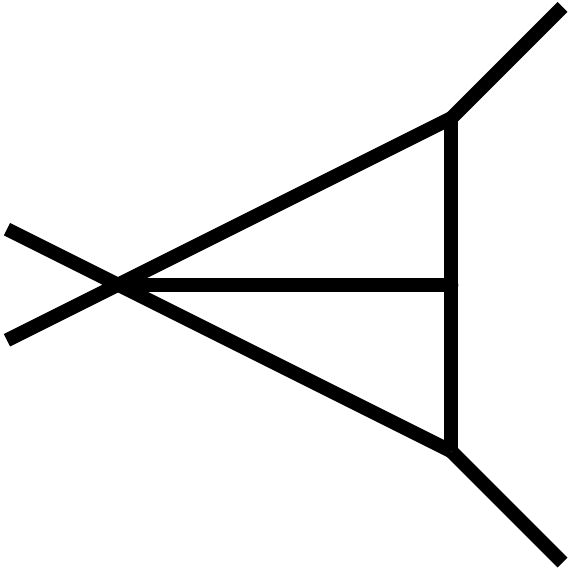}};
    \node at
    (2.4,1.4){\includegraphics[scale=0.22]{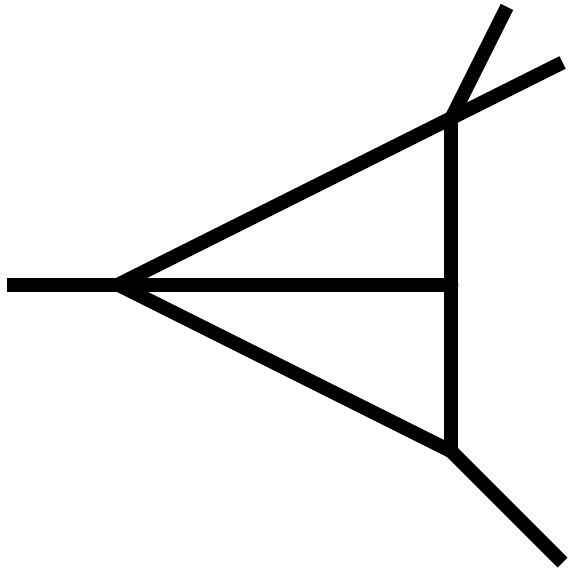}};
    \node at
    (3.2,1.4){\includegraphics[scale=0.22]{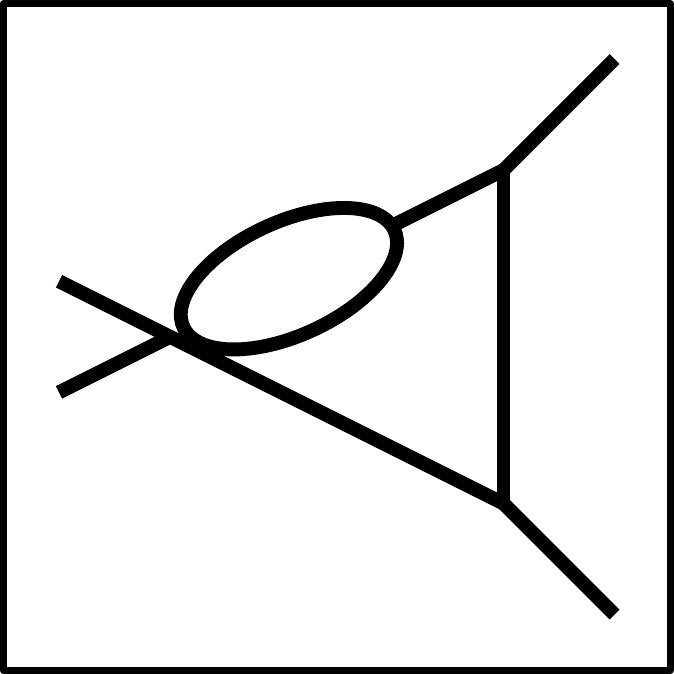}};
    \node at
    (4,1.4){\includegraphics[scale=0.22]{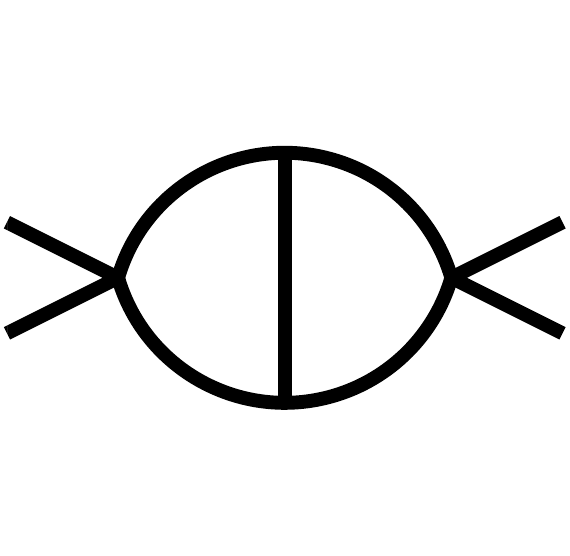}};
    \node at (4.8,1.4){\includegraphics[scale=0.22]{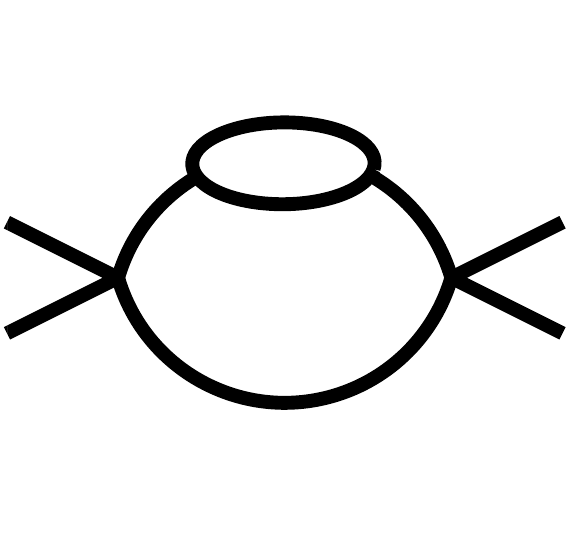}};
    \node at (5.8,1.4){\includegraphics[scale=0.22,angle=180]{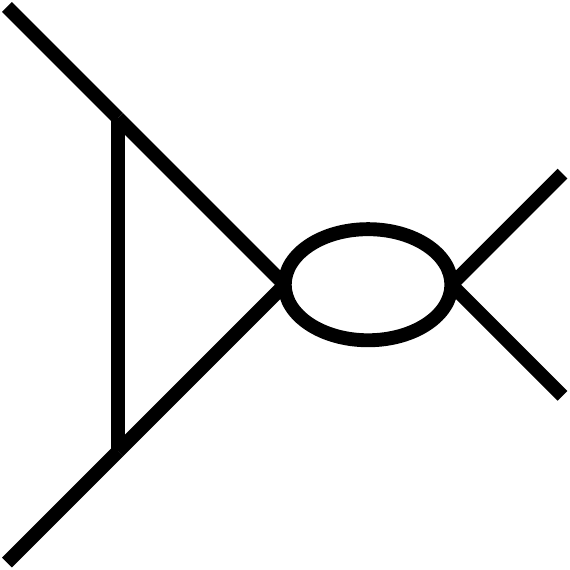}};
    \node at
    (6.5,1.4){\includegraphics[scale=0.22]{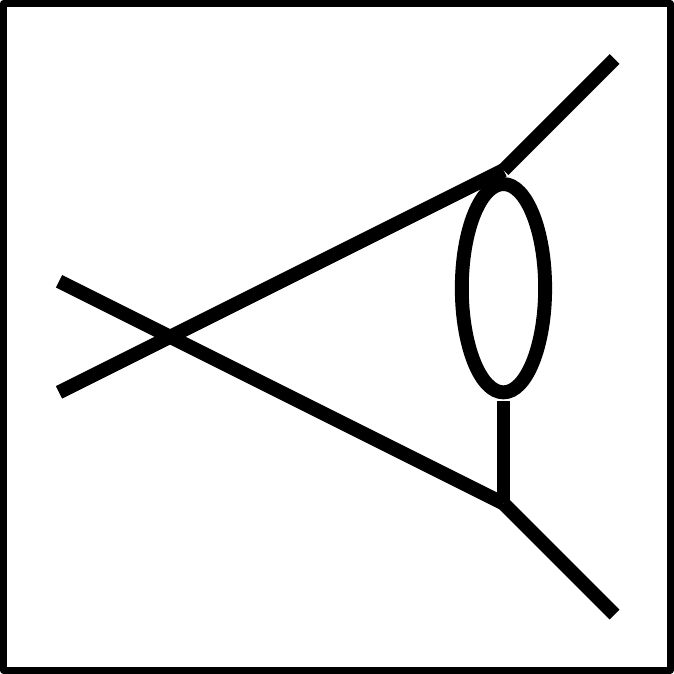}}; 
    \node at
    (1.75,.5){\includegraphics[scale=0.22]{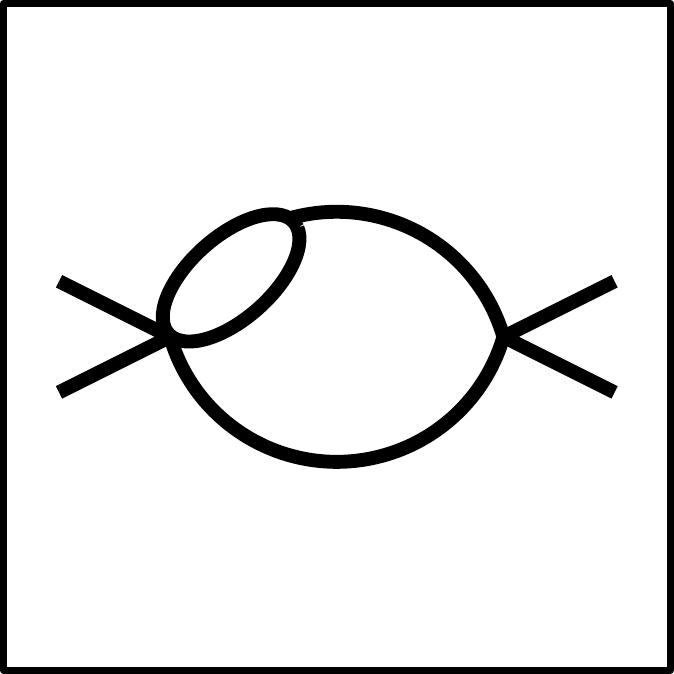}};
    \node at
    (3.35,.5){\includegraphics[scale=0.22]{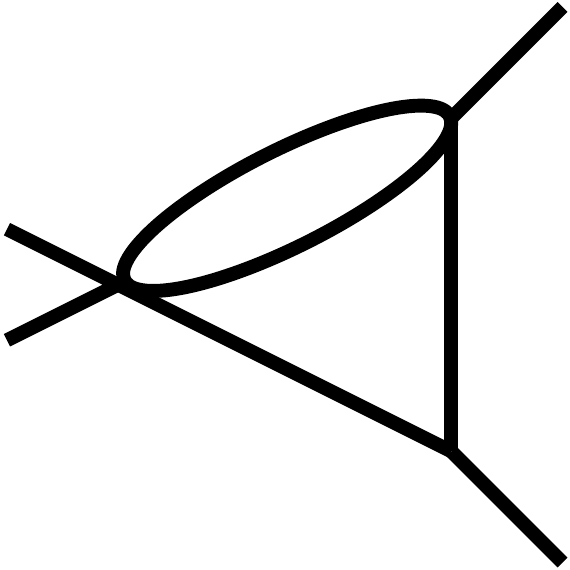}}; 
    \node at
    (5.8,.5){\includegraphics[scale=0.22]{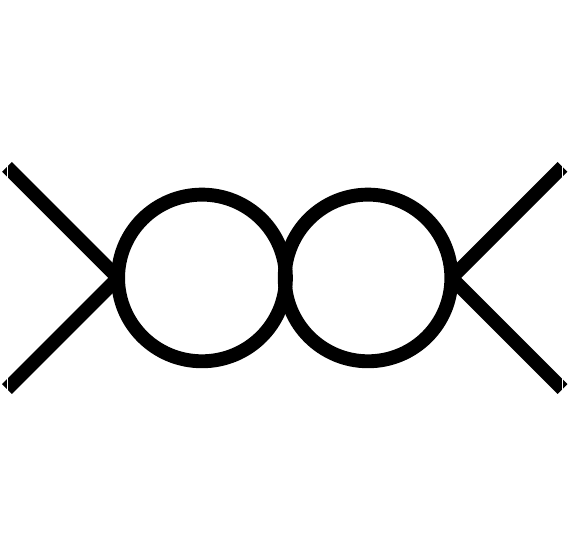}};
    \node at
    (6.5,.5){\includegraphics[scale=0.22]{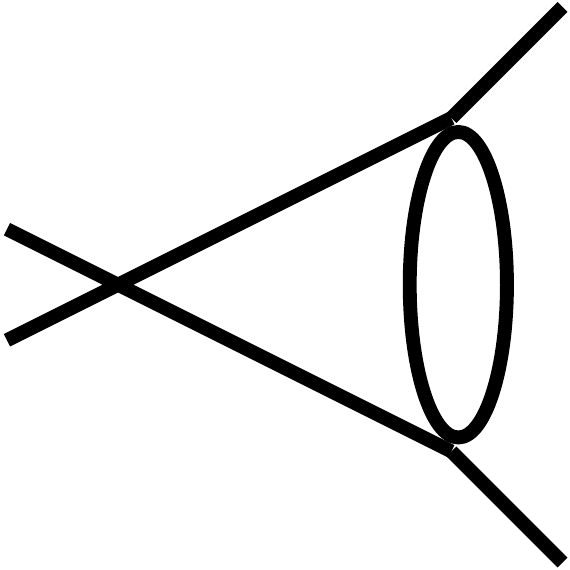}}; 
    \node at (2.55,-0.2){\includegraphics[scale=0.27]{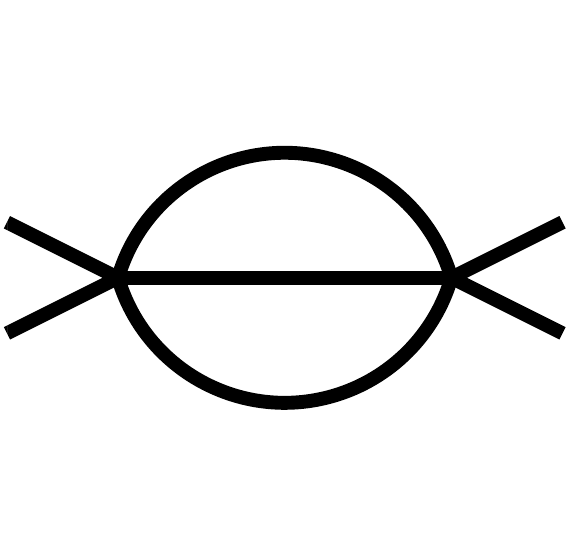}};
\end{tikzpicture} \caption{The planar $\Delta$ hierarchy in a 2 $\to$ 2 amplitude.
    Only topologically
    inequivalent diagrams are shown.
    The boxed diagrams do not belong to the cut hierarchy.
    The diagrams to the left of the dashed
    line are the members of the sunrise (cut) hierarchy.}
\label{fig:diagSunrise}
\end{figure}

The required tree amplitudes
can also be indexed by the diagrams $\Gamma$. Let $T_\Gamma$ denote the set
of tree amplitudes constructed in the following way:
to each $n$-point vertex of $\Gamma$, one
associates an $n$-point tree amplitude 
evaluated on the on-shell
momenta $\ell_l^\Gamma$,
with matched quantum numbers on internal lines. 
Notice that not all diagrams $\Gamma$ give rise to a well defined
product of tree amplitudes. An example of
this is displayed in \fig{fig:DProptopologies}.
The tree amplitude associated to the four-point vertex on
the top right corner of diagram (b) behaves as $1/\rho$ and thus is divergent
and ill-defined in the on-shell phase space of the diagram (which sets $\rho=0$).
We will denote the subset of all
diagrams which give rise to a well
defined product of tree amplitudes by
\begin{equation}
\Delta'\subseteq\Delta \,. \label{eq:Deltap}
\end{equation}
Note that $\Delta'$ inherits the notion of hierarchy from $\Delta$. The hierarchies
in $\Delta'$ are referred to as {\it cut hierarchies}. It is a general feature that
$\Delta\neq\Delta'$ whenever one of the diagrams
$\Gamma\in\Delta$ has a propagator
structure with multiple copies of a given propagator, like diagram (a)
in \fig{fig:DProptopologies}.
In \fig{fig:diagSunrise}, to the left of the dashed line, we show the
{\it sunrise} hierarchy for a massless 2~$\to$~2 amplitude. The diagrams not belonging
to the corresponding cut hierarchy are drawn inside a box.

Generalized unitarity builds on the observation that, in the limit $\ell_l\to\ell_l^\Gamma$,
we have
\begin{eqnarray} 
    \lim_{ \ell_l\to\ell_l^\Gamma  }\,{\cal A}(\ell_l) &=&
    \frac{1}{\prod_{k\in P_\Gamma} \rho_k} \,
    \left(R(\Gamma, \ell_l^\Gamma)
     +{\cal O}(\rho_{k\in P_\Gamma})
    \right) \quad\mbox{for each}\quad  \Gamma\in\Delta'\,, \label{eq:cut}
\end{eqnarray} 
and that in this limit $R(\Gamma, \ell_l^\Gamma)$ is given as a product of trees,
\begin{eqnarray} 
    R(\Gamma, \ell_l^\Gamma) &=& \sum_{\rm states} \prod_{k\in T_\Gamma} {\cal A}^{\rm
    tree}_k(\ell_l^\Gamma) \quad\mbox{for each}\quad  \Gamma\in\Delta' \,.
    \label{eq:R}
\end{eqnarray}
We stress the fact that $R(\Gamma, \ell_l^\Gamma)$ is only defined on the
on-shell phase space of $\Gamma$.
Naively, in generalized unitarity, one expects one equation (\ref{eq:cut})
for each diagram, such that each numerator $N(\Gamma,\ell)$,
as defined in \eqn{eq:N},
is associated to its individual on-shell limit. However, starting at two-loops,
$\Delta'\neq \Delta$ and \eqn{eq:R} is well defined only for
the subset of diagrams $\Gamma\in\Delta'\subset \Delta$.
We are thus left with less equations to determine
the coefficients in \eqn{eq:AL}.
In the following subsection we outline our algorithm to overcome this
issue, and in \sect{sec:applications} we apply
it in the context of concrete two-loop examples.

\subsection{Leading and subleading poles in generalized unitarity}

The diagrams in $\Delta'$ correspond to leading poles of the amplitude
in the on-shell limit, while those in $\Delta\setminus\Delta'$ correspond to subleading poles.
We start by reviewing the standard approach to deal with leading poles.

\subsubsection{Extracting leading poles}\label{sec:leadingPolesGen}

Consider for simplicity the integrand of a maximal diagram $\Gamma$, that is a
configuration in which $\Gamma$ contains the maximal number of edges
required for the amplitude $\cal{A}$.
When working in $D$ dimensions and with external momenta defined in four
dimensions,
the maximum number of edges is bounded for a two-loop amplitude with $n$
external particles by ${\rm min}(n+3,11)$.
Furthermore, each subloop can contain at most 6 edges (7 if a doubled propagator is present).
On the on-shell phase space of $\Gamma$, we get from \eqn{eq:cut} that
\begin{equation} 
N(\Gamma,\ell_l^{\Gamma}) = R(\Gamma,\ell_l^\Gamma)\ .  \label{eq:maxcut} 
\end{equation}
Through unitarity, we can directly compute $N(\Gamma,\ell_l^{\Gamma})$ as a
product of trees, see \eqn{eq:R}.
Using \eqn{eq:N}, we can then extract the corresponding set
$\{c_{\Gamma,i}\}$ of integrand coefficients by sampling 
\eqn{eq:maxcut} over enough points in the on-shell phase space $\ell_l^{\Gamma}$.
We thus obtain $N(\Gamma,\ell_l)$
for generic $\ell_l$.

Consider now a next-to-maximal integrand, that is an integrand with one less
propagator than a maximal one. We can still use a relation similar to
\eqn{eq:maxcut}, taking care of subtracting contributions coming from integrands
with more propagators. 
For concreteness, let $\Gamma_\text{NM}$ be a next-to-maximal diagram.
We denote by $\Gamma_{\text{NM},k}$ the ancestor
of $\Gamma_\text{NM}$ which has the same propagators
as $\Gamma_\text{NM}$ plus an extra one, $1/\rho_k$.
At this stage we assume that $k\notin P_{\Gamma_{\rm NM}}$
for any~$k$.
On the on-shell phase space of $\Gamma_\text{NM}$,
unitarity ensures 
\begin{align}\label{eq:nmcut} 
    {N}^{}(\Gamma_\text{NM},\ell_l^{\Gamma_\text{NM}})
    =R(\Gamma_\text{NM},\ell_l^{\Gamma_\text{NM}}) - 
    \sum_k \frac{1}{\rho_k(\ell_l^{\Gamma_\text{NM}})}
    {N}(\Gamma_{\text{NM},k},\ell_l^{\Gamma_\text{NM}})\ ,
\end{align} 
in which the inverse propagators $\rho_k$ are evaluated on the momenta
$\ell_l^{\Gamma_\text{NM}}$.  The coefficients $\{c_{\Gamma_\text{NM},i}\}$
in the numerator $N(\Gamma_\text{NM},\ell_l)$ are determined from a
linear system of equations obtained
from \eqns{eq:N}{eq:nmcut}.

In the absence of subleading poles, we can iterate this procedure. 
A systematic extraction of all integrand coefficients of a given
amplitude, see \eqn{eq:AL}, can then be carried out ``level by level'', from
the maximal integrands to the minimal ones.

\subsubsection{Extracting subleading poles}

Let us now consider the case where subleading singularities are present
and develop an algorithm to extract their contributions. Let
$\Gamma_p$ and $\Gamma_d$ be a pair of parent-daughter diagrams, $\Gamma_p>\Gamma_d$, such that
the inverse propagator
$\rho_s$ appears both in $\Gamma_p$ and $\Gamma_d$, but to a higher power
in $\Gamma_p$ than in~$\Gamma_d$. At two loops, it is sufficient to assume that
it is squared in $\Gamma_p$.
According to the definition of \sect{sec:genUni},
$\Gamma_p\in\Delta'$ but $\Gamma_d\notin\Delta'$. 

Then, the on-shell phase space defined by the two diagrams is the same,
schematically $\ell_l^{\Gamma_p}=\ell_l^{\Gamma_d}$,
and we thus have both a leading (from $\Gamma_p$) and sub-leading pole
contribution (from $\Gamma_{d}$) in the
on-shell limit of the integrand,
\begin{eqnarray} 
\lim_{ \ell_l  \to \ell_l^{\Gamma_d}  }\,{\cal A}(\ell_l) &=&
\frac{1}{\prod_{k\in P_{\Gamma_p}} \rho_k} \, \left( 
R(\Gamma_p,\ell_l^{\Gamma_p})+\rho_s
R(\Gamma_d,\ell_l^{\Gamma_d})
+{\cal O}(\rho_{k\in P_{\Gamma_d}}) \right)\,,
\label{eq:partfr}
\end{eqnarray} 
where $R(\Gamma_d,\ell_l^{\Gamma_d})$, for $\Gamma_d\notin\Delta'$, has been
implicitly defined as the subleading term in the $\ell_l  \to \ell_l^{\Gamma_d}$
limit. We stress again, that this definition applies only on the on-shell phase
space of $\Gamma_d$.

The term $R(\Gamma_p,\ell_l^{\Gamma_p})$ is the leading term
in the $\ell_l  \to \ell_l^{\Gamma_d}$ limit, and is obtained from
a product of trees as in \eqn{eq:R}. From it we can determine the associated
numerator ${N}(\Gamma_{p},\ell_l)$ in the standard way discussed
in the previous section. In contrast, 
an equivalent expression for $R(\Gamma_d,\ell_l^{\Gamma_d})$ is not known
and the determination of the corresponding numerator
${N}(\Gamma_{d},\ell_l)$ must proceed differently.

We carry on as follows. We go down the $\Delta$ hierarchy
until we find a diagram that has all the same propagators as $\Gamma_d$ except 
$1/\rho_s$. In our case it is sufficient to consider a corresponding
daughter diagram of $\Gamma_d$ (in turn, a granddaughter of $\Gamma_p$).
Let $\Gamma'$ be such a diagram,  $s\notin P_{\Gamma'}$.
Assuming that the only subleading poles are associated with the propagator $s$
(that is, $\Gamma'\in\Delta'$ by construction),
the factorization limit of the amplitude as
$\ell_l\to\ell_l^{\Gamma'}$ leads to
\begin{equation}\label{eq:cutEq}
    R(\Gamma',\ell_l^{\Gamma'})
    =
    N(\Gamma',\ell_l^{\Gamma'})
    +\sum_{\substack{\Gamma\,\in\,\Delta \\ \Gamma>\Gamma'}}
    \frac{N(\Gamma,\ell_l^{\Gamma'})}
    {\prod_{k\in\,P_{\Gamma}\setminus P_{\Gamma'}} \rho_k(\ell_l^{\Gamma'})}\,,
\end{equation}
with $R(\Gamma',\ell_l^{\Gamma'})$ given by a product of trees, see \eqn{eq:R}.
We call \eqn{eq:cutEq} the {\it cut equation}, and there exists one cut equation
for each element of $\Delta'$.
The sum over diagrams $\Gamma$ runs over all ancestors of $\Gamma'$.  In a standard
unitarity approach and in the absence of subleading poles, all numerators
$N(\Gamma,\ell_l^{\Gamma'})$ for $\Gamma>\Gamma'$ will have been
determined previously from their own cut equations.
We would thus use \eqn{eq:cutEq} to determine
the numerator $N(\Gamma',\ell_l^{\Gamma'})$, see e.g.~\eqn{eq:nmcut} where
this is done explicitly. 
In the presence of subleading poles,
some numerators $N(\Gamma,\ell_l^{\Gamma'})$ cannot
be determined from an associated cut equation,
as already discussed below \eqn{eq:R}. We thus separate the terms of the sum in \eqn{eq:cutEq}
into two sets: those in $\widetilde \Delta$ whose numerators are still unknown, and those
in $\Delta\setminus \widetilde\Delta$ which have already been determined. We then rewrite
\eqn{eq:cutEq} as:
\begin{equation}\label{eq:mainResult}
    N(\Gamma',\ell_l^{\Gamma'})+
    \sum_{\substack{\Gamma\,\in\,\widetilde\Delta \\ \Gamma>\Gamma'}}
    \frac{N(\Gamma,\ell_l^{\Gamma'})}{\prod_{k\in\,P_{\Gamma}\setminus P_{\Gamma'}} \rho_k(\ell_l^{\Gamma'})}
    =R(\Gamma',\ell_l^{\Gamma'})
    -\sum_{\substack{\Gamma\,\in\,\Delta\setminus \widetilde\Delta\\ \Gamma>\Gamma'}}
    \frac{N(\Gamma,\ell_l^{\Gamma'})}
    {\prod_{k\in\,P_{\Gamma}\setminus P_{\Gamma'}} \rho_k(\ell_l^{\Gamma'})}\,.
\end{equation}
In this expression, all numerator terms on the right-hand side can be extracted from
the standard generalized unitarity approach outlined in \sect{sec:leadingPolesGen}.
All numerator terms on the left-hand side are still to be determined, and we will do so
for all at once. More precisely, we sample \eqn{eq:mainResult} over
enough points of the on-shell phase space $\ell_l^{\Gamma'}$ to build
a system of equations big enough to determine all coefficient functions
$\{c_{\Gamma',i}\}$ and $\{c_{\Gamma,i}\}$ for all $\Gamma>\Gamma'$,
$\Gamma\in\widetilde\Delta$.

We note that the algorithm that we have proposed relies only on the unitarity
of the theory under consideration. In particular, it extends trivially to any
loop order and is entirely process independent. Indeed, while for generic multi-loop
amplitudes the structure of subleading poles is in general much
richer, with for example more than one subleading term in on-shell limits like
in~\eqn{eq:partfr}, mixed subleading poles associated to different
propagators, and non-planar configurations, our algorithm still allows to
find enough suitable cut equations to solve for all unknown numerators as
in~\eqn{eq:mainResult}.

\section{Applications to four-point gluon amplitudes}\label{sec:applications}

In this section we apply the algorithm introduced in the previous section to extract
coefficients of ancestor diagrams on the phase space of its descendants in the context
of one- and two-loop examples. The one-loop example is included as a simple
illustration of our algorithm, to show that
it can also be used in the absence of subleading poles.
In the two-loop example, we apply it to a case with subleading poles
where a standard generalized unitarity approach would not be enough. Finally,
we discuss the implementation of our approach in a numerical framework
and the checks that we have performed on its applicability.

\subsection{Box coefficients from the triangle phase-space}

We compute box and triangle coefficients
from triple cuts only.  The system
of equations which arises is less diagonal than a standard one-loop approach,
but nonetheless tractable. Consider the expression for a triangle cut of a
four-point amplitude at one-loop, a specific example of the next-to-maximal
case described in \eqn{eq:nmcut},
\begin{equation} 
N \left(\eqnDiag{\includegraphics[scale=0.1]{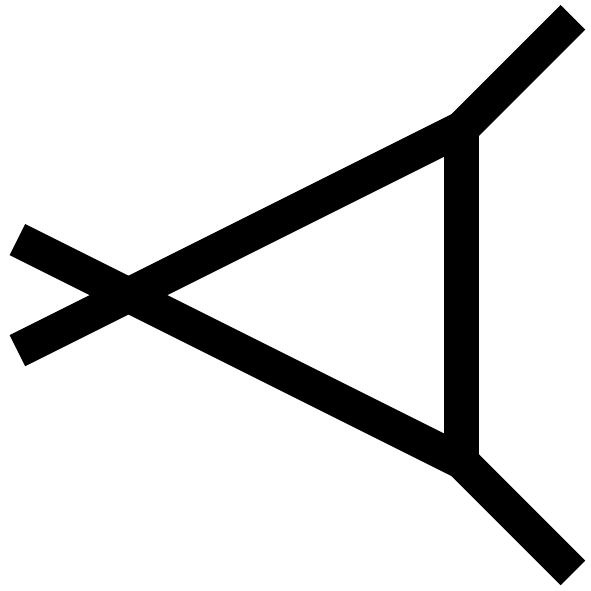}},\ell^{\rm tri
}\right)
= R \left(\eqnDiag{\includegraphics[scale=0.1]{diagrams/Triangle}},\ell^{\rm
tri }\right) - \frac{1}{\rho} N
\left(\eqnDiag{\includegraphics[scale=0.1]{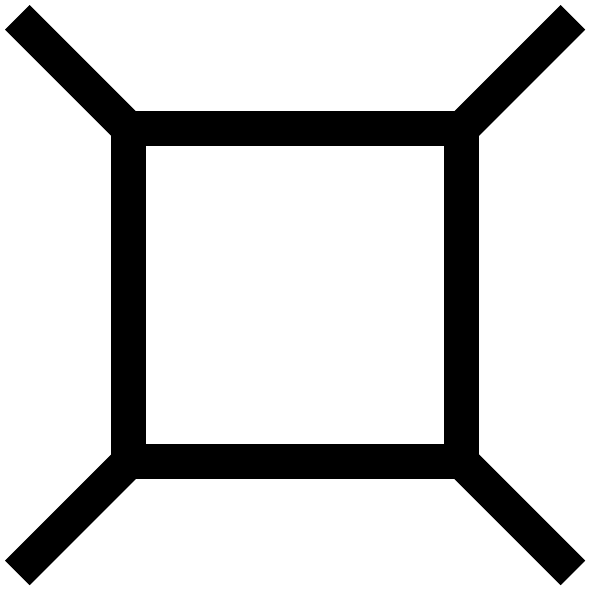}},\ell^{\rm tri
}\right)\ , \label{eq:ModifiedBoxSubtraction}
\end{equation}
where $\rho$ is the inverse propagator that was pinched to obtain the triangle
diagram from the box diagram. The propagator is evaluated on the on-shell
momentum  $\ell^{\rm tri}$. We have used a pictorial representation to show the
contributing diagrams. This is the standard cut equation
for the triangle and the (parent-)box cuts, however, we assume that the box
coefficient functions
have not yet been determined, i.e.,
the box diagrams belong to $\widetilde \Delta$ in \eqn{eq:mainResult}.
To proceed we insert the corresponding expressions for the box and triangle
numerator functions, see \eqn{eq:N}, leading to
\begin{equation}\label{eq:cutlessTriBox}
\sum_{i=1}^{m} c_{{\rm tri},i} m_{{\rm tri},i} (\ell^{\rm tri})
+ \frac{1}{\rho} \sum_{i=1}^{n} c_{{\rm box},i}\, m_{{\rm box},i} (\ell^{\rm
tri})
=
R \left(\eqnDiag{\includegraphics[scale=0.1]{diagrams/Triangle}}, \ell^{\rm
tri}\right) \,,
\end{equation}
in which the right-hand side is given by a product of trees. 
The triangle and box coefficients are written as $c_{{\rm tri},i}$ and
$c_{{\rm box},i}$, respectively.  The associated numerator insertions are
denoted by $m_{{\rm tri},i}(\ell)$ and  $m_{{\rm box},i}(\ell)$.  
The number of master and surface integrands of the triangle and box
diagrams have been denoted by $m$ and $n$, respectively. Compared to the
notation in \eqn{eq:AL}, $m$
is the number of elements of $M_{\rm tri}\cup S_{\rm tri}$ and $n$
the number of elements of $M_{\rm box}\cup S_{\rm box}$.
All the triangle and box coefficients are then found by sampling the
triangle cut over $n+m$ momenta on the on-shell phase space $\ell^{\rm tri}$
and then solving for $c_{{\rm box},i}$ and
$c_{{\rm tri},i}$ by a linear regression.
This one-step approach requires solving a single large linear
system of equations, compared to two smaller ones when solving first for box
coefficients and subsequently for the triangle coefficients.

\subsection{The bubble-box hierarchy at two loops}

As an example of the application of our algorithm in the presence of subleading
poles, we consider a 2 $\to$ 2 amplitude in massless QCD. The maximal level
diagrams have seven propagators, and the minimal diagrams
are the sunrise diagrams with three propagators (see \fig{fig:diagSunrise}).

%
\begin{figure}[ht]
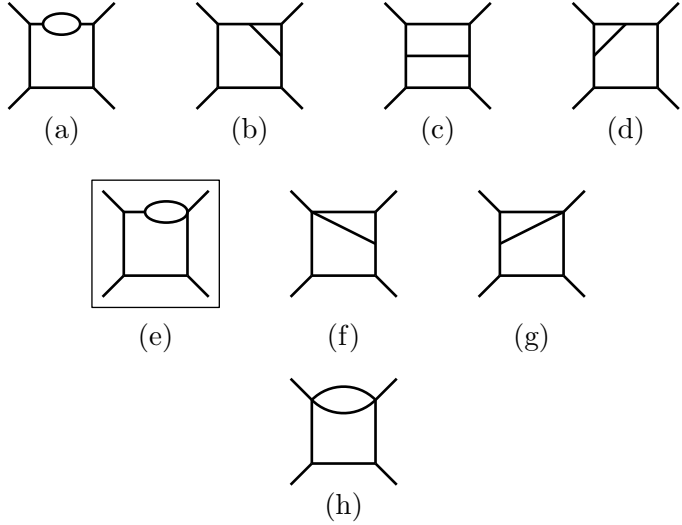
 \begin{tikzpicture}[scale=2.5]
        \node at (0,2){\includegraphics[scale=0.25]{diagrams/HexaBubbleThin}};
        \node at (0,1.6){(a)}; \node at
        (1,2){\includegraphics[scale=0.25]{diagrams/PentaTriangleThin}}; \node
        at (1,1.6){(b)}; \node at
        (2,2){\includegraphics[scale=0.25]{diagrams/DoubleBoxThin}}; \node at
        (2,1.6){(c)}; \node at
        (3,2){\reflectbox{\includegraphics[scale=0.25]{diagrams/PentaTriangleThin}}};
        \node at (3,1.6){(d)};
        \node at
        (0.5,1){\includegraphics[scale=0.25]{diagrams/BubblePentagonSemiGenericBoxedThin}};
        \node at (0.5,0.5){(e)}; \node at
        (1.5,1){\includegraphics[scale=0.25]{diagrams/BoxTriangleThin}}; \node
        at (1.5,0.5){(f)}; \node at
        (2.5,1){\reflectbox{\includegraphics[scale=0.25]{diagrams/BoxTriangleThin}}};
        \node at (2.5,0.5){(g)};
        \node at
        (1.5,0){\includegraphics[scale=0.25]{diagrams/BubbleBoxGenericThin}};
        \node at (1.5,-0.4){(h)}; \end{tikzpicture}
\caption{The planar bubble-box hierarchy. The maximal diagrams are (a)-(d),
next-to maximal are the (e)-(g) and at the bottom we find the bubble-box
diagram (h).}
\label{fig:diagDprop} \end{figure}

Consider now the {\it bubble-box} hierarchy shown in \fig{fig:diagDprop}. All diagrams that appear above it are
associated with factorization limits of (h).  Six out of the
seven ancestor diagrams shown have associated products of trees
and their integrand
coefficients can be directly extracted.  Diagram (e), on the other hand,
represents subleading pole contributions to the doubled-propagator
diagram (a). 

As an aside, we note that for all numerators in \fig{fig:diagDprop}, apart from the 
double-box (c) and the  bubble-box (h), the integrand function space is
spanned entirely by surface terms. Two master integrals are associated with (c)
and one with (h). With this in hand, we solve the cut hierarchy to obtain the coefficients.

Consider first the integrands associated with the maximal diagrams (a)-(d). 
For example, for the numerator of (c), we have:
\begin{eqnarray}
N\left(\eqnDiag{\includegraphics[scale=0.1]{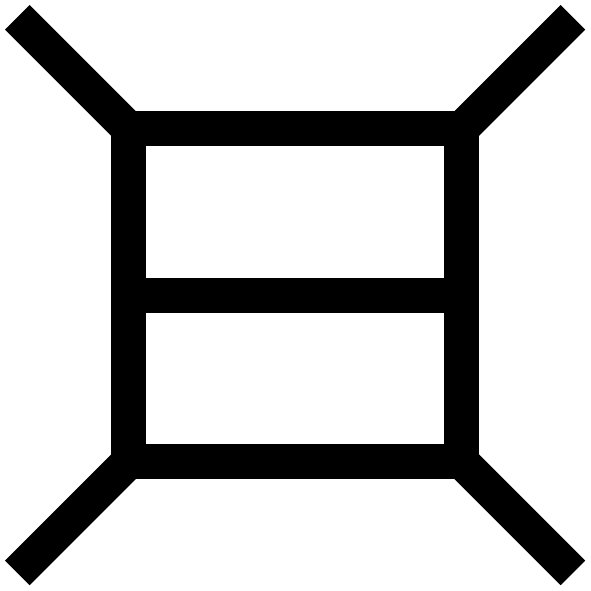}},\ell_l^{\rm c}\right)&=&
R\left(\eqnDiag{\includegraphics[scale=0.1]{diagrams/DoubleBoxSmall.pdf}},\ell_l^{\rm c}\right)
\,,  \label{eq:DBeq} 
\end{eqnarray} 
where the on-shell momenta of diagram (c) are denoted by $\ell_l^c$. We have
used a pictorial representation to denote the corresponding diagram. Analogous
equations hold for the diagrams (a), (b) and (d). 
In practice, for each maximal diagram $\Gamma$ one generates a linear set
of equations by inserting sufficiently many on-shell momentum values for
$\ell_l^\Gamma$, and solves for the integrand coefficients in $N(\Gamma,\ell_l)$.

We move then to the
numerators of the two next-to-maximal diagrams, (f) and (g). These are found
by their corresponding cut equations, as in \eqn{eq:nmcut}. For example, 
the numerator for the {\it box-triangle} diagram (f), fulfills 
\begin{eqnarray}
    N\left(\eqnDiag{\includegraphics[scale=0.1]{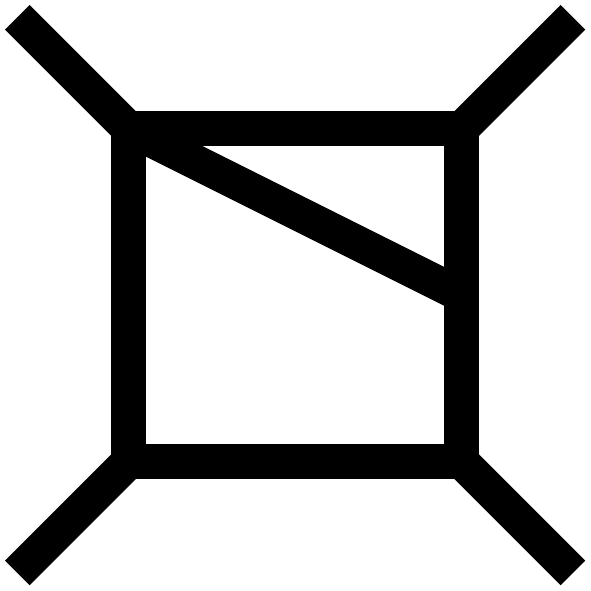}},\ell_l^{\rm f}\right)=
    R\left(\eqnDiag{\includegraphics[scale=0.1]{diagrams/BoxTriangle1Small.pdf}},\ell_l^{\rm f}\right)
    - \frac{1}{\rho_{\rm fb}}
    N\left(\eqnDiag{\includegraphics[scale=0.1]{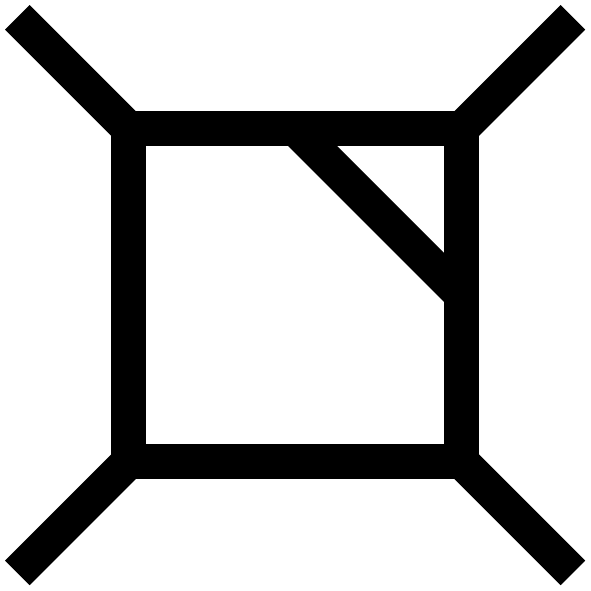}},\ell_1^{\rm f}\right)
    - \frac{1}{\rho_{\rm fc}}  N\left(
    \eqnDiag{\includegraphics[scale=0.1]{diagrams/DoubleBoxSmall.pdf}},\ell_l^{\rm f}\right)\
    ,\nonumber 
\end{eqnarray}
where $\rho_{\rm fb}$ and $\rho_{\rm fc}$ denote the propagators that are pinched to
obtain diagram (f) from (b) and diagram (f) from (c), respectively (see
\fig{fig:diagDprop}). We denote the on-shell momenta associated to
diagram (f) by $\ell_l^{\rm f}$. The integrand
corresponding to diagram (g) is treated in the same manner.

Finally, we proceed to solve for the coefficients associated to diagrams (e)
and (h), which involves the extraction of subleading poles. 
Setting  $\widetilde\Delta=\{(e)\}$ in \eqn{eq:mainResult} we obtain
\begin{align} 
    N\left(
    \eqnDiag{\includegraphics[scale=0.1]{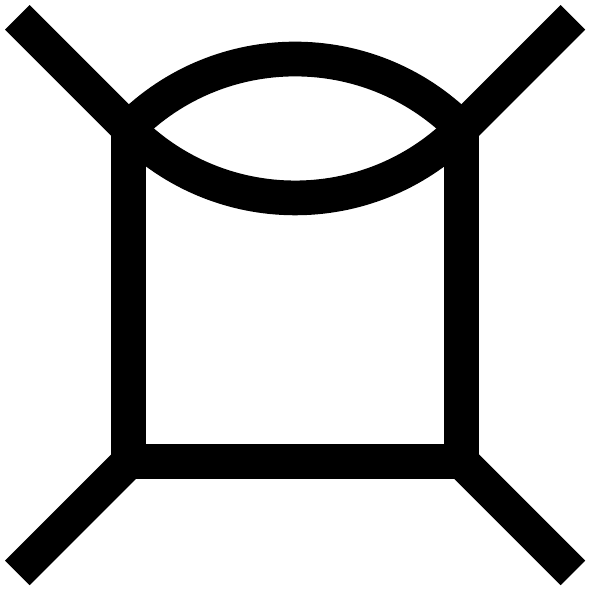}},\ell_l^{\rm h}
    \right) +& \frac{1}{\rho_{\rm he}}{N}\left(
    \eqnDiag{\includegraphics[scale=0.1]{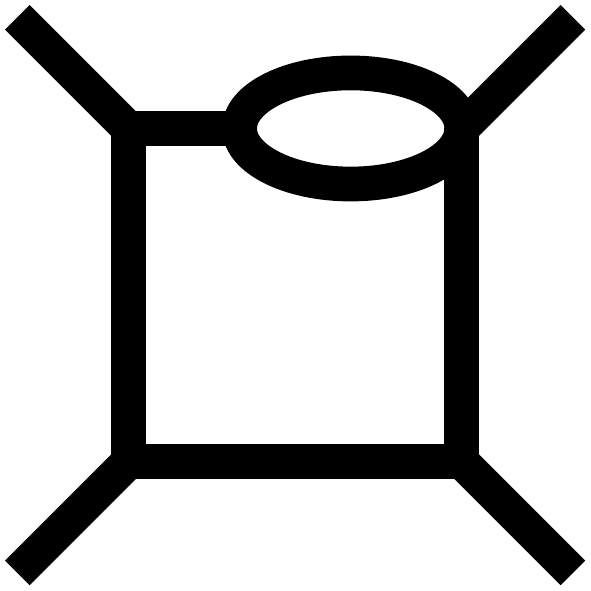}},\ell_l^{\rm h}
    \right)=\notag\\
    &\quad   R\left(
    \eqnDiag{\includegraphics[scale=0.1]{diagrams/BubbleBoxGeneric.pdf}}, \ell_l^{\rm h} \right) 
    - \frac{1}{\rho_{\rm hf}}  {N}\left(
    \eqnDiag{\includegraphics[scale=0.1]{diagrams/BoxTriangle1Small.pdf}},\ell_l^{\rm h} \right)
    - \frac{1}{\rho_{\rm hg}}  {N}\left(
    \eqnDiag{\includegraphics[scale=0.1]{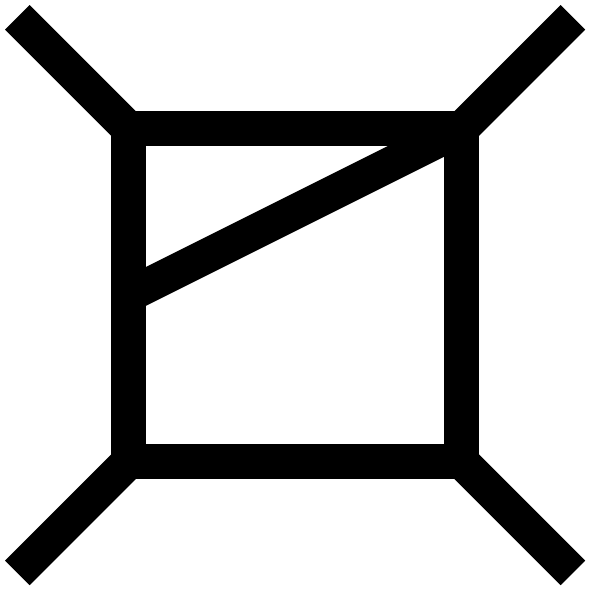}},\ell_l^{\rm h} \right)
    - \frac{1}{(\rho_{\rm he})^2} {N}\left(
    \eqnDiag{\includegraphics[scale=0.1]{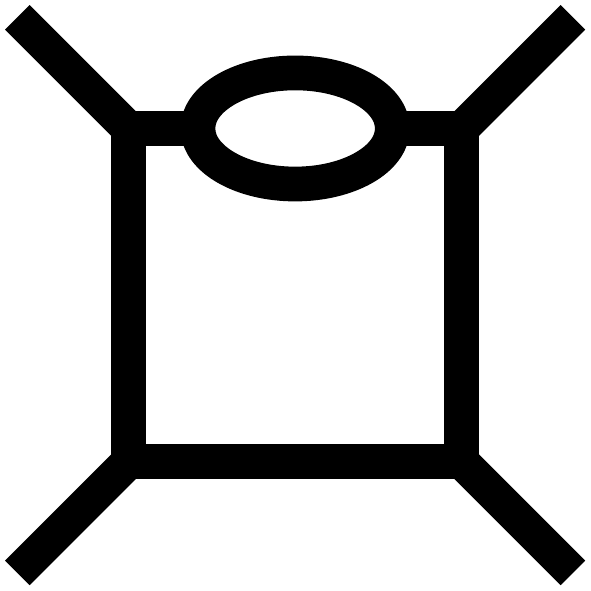}},\ell_l^{\rm h} \right)
    \notag \\ &\quad - \frac{1}{\rho_{\rm hf}\rho_{\rm fb}}  {N}\left(
    \eqnDiag{\includegraphics[scale=0.1]{diagrams/PentaTriangle1Small.pdf}},\ell_l^{\rm h} \right)
    - \frac{1}{\rho_{\rm hf}\rho_{\rm fc}}  { N}\left(
    \eqnDiag{\includegraphics[scale=0.1]{diagrams/DoubleBoxSmall.pdf}},\ell_l^{\rm h} \right) -
    \frac{1}{\rho_{\rm hg}\rho_{\rm gd}}  {N}\left(
    \eqnDiag{\includegraphics[scale=0.1]{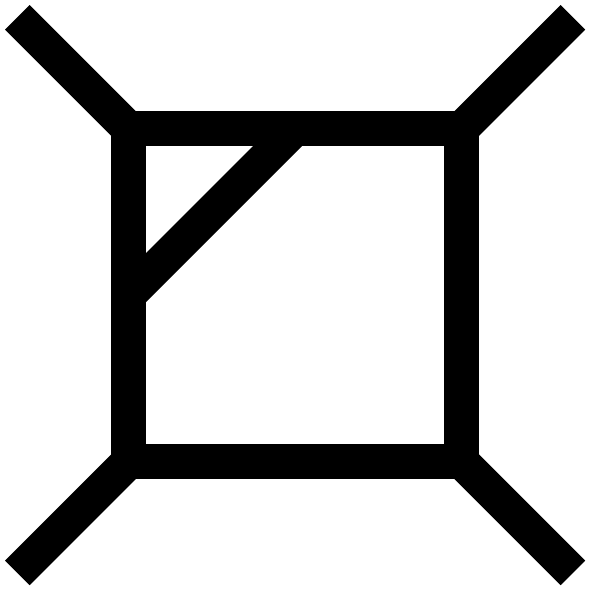}}, \ell_l^{\rm h} \right)\
    , \label{eq:BBeq} 
\end{align} 
where the inverse propagators $\rho_{ij}$ denote the  propagator pinched to
obtain diagram ($i$) from diagram ($j$), with the indices ($i$) and ($j$) 
corresponding to pairs of daughter-parent diagrams in \fig{fig:diagDprop}. 
Each numerator ${N}(\Gamma,\ell_l^{\rm h})$ is written in terms of its basis of
integrands as in \eqn{eq:N}. At this stage, the terms on the right-hand side of \eqn{eq:BBeq}
are known, and we can directly solve for the coefficients associated to the
integrand 
${N}\left(
\eqnDiag{\includegraphics[scale=0.1]{diagrams/BubbleBoxGeneric.pdf}},\ell_l^{\rm
h}  \right)$
as well as to the subleading-pole integrand 
${N}\left(
\eqnDiag{\includegraphics[scale=0.1]{diagrams/BubblePentagonSemiGeneric.pdf}},\ell_l^{\rm
h} \right)$
by sampling over enough on-shell loop momentum configurations.

\subsection{Numerical implementation and checks} \label{sec:checks}

We discuss in this section the checks that we have performed on the
applicability of our algorithm in
a numerical unitarity framework.
We have numerically reproduced 
the analytic results of~\cite{fourgluons}
for all the master coefficient functions in all planar two-loop four-gluon
helicity amplitudes.
This was achieved through the following steps:
\begin{itemize} 
\item We have implemented $D$-dimensional off-shell recursions~\cite{BGrec} to
    compute all required trees and (one- and) two-loop cuts for generic
    computations in $D$-dimensional numerical unitarity. Our implementation
    allows general values of the $D_s$ parameter,
    the dimensionality of the spin space for the loop particles. The
    numerical trees and one-loop cuts were cross checked against the
    \BlackHat{} library~\cite{BlackHat} in four dimensions. We have also cross checked our
    two-loop cuts against results obtained using an independent implementation
    of the gluon amplitudes in six dimensions. Systematic checks of Ward
    identities and factorization limits of the trees and cuts have been
    performed.
\item Two independent implementations for the construction of the $\Delta$
    hierarchies and the associated subtraction structures have been produced, one
    based on planar configurations and another following the color decomposition
    of~\cite{Ochirov:2016ewn}. Both have been cross checked, and independently shown
    to correctly produce  subtraction terms for diagrams in $\Delta'$.
\item In order to compare against known results for two-loop four-point gluon
    amplitudes, we have produced a set of master-surface integrand 
    decompositions as in \eqn{eq:AL} for all the diagrams $\Gamma$ in those
    amplitudes. The decompositions were produced along
    the lines of ref.~\cite{NumUnitarity2}.
    We cross-checked that the surface integrands we constructed integrate
    to zero with the generator of \ibp{} relations FIRE~\cite{Smirnov:2014hma}.
    We have also shown that the integrand decompositions fully span their
    corresponding integrand spaces by comparing them to an alternative parametrization
    in terms of tensor insertions \cite{tensorBasis}.
\item With all the tools described above, we were able to numerically compute all
    integrand coefficients that contribute to the planar two-loop four-point gluon
    amplitude, see \fig{fig:diagSunrise}. 
    We have validated the values of the master integral
    coefficients with the known
    analytic results~\cite{fourgluons} for all helicity configurations.
\end{itemize}

Most checks have been performed on a small set of phase-space points and we
delay any systematic efficiency and stability checks of our implementation to
future work.  Nevertheless it is worth mentioning that we find that for example the master
coefficient functions of the bubble-box hierarchy in \fig{fig:diagDprop} can be
extracted in less that 100$ms$ on generic phase-space points.  Typically we find
that they agree with numerical values obtained from the analytic expressions
\cite{fourgluons} to an accuracy of better than 10 digits.  
This study has been performed for fixed values of the dimensional parameters
$D$ and $D_s$ using only double-precision arithmetics.

\section{Conclusions} \label{sec:outlook}

In this article we have presented an algorithm for extracting subleading-pole
contributions in two-loop scattering amplitudes through numerical unitarity
techniques, which are required to obtain the full amplitude.
Subleading-pole integrand coefficients can be obtained by solving
linear systems of equations built up from cut equations of associated descendant
diagrams. 
The algorithm is process independent and can be naturally generalized to
multi-loop amplitudes. 
We have performed a number of consistency checks, in which multiple
subleading-pole contributions have been extracted at several levels of a two-loop
cut hierarchy. 
In addition, we computed two-loop master
coefficients through numerical unitarity, and have confirmed our
results by comparing to available analytic results for four-point two-loop gluon
amplitudes.
Although in numerical unitarity one performs calculations in fixed
dimensions $D$ and $D_s$, the regressions to general values of those
parameters can be achieved by the observation that functionally they appear as
rational or polynomial functions. We have reconstructed the full $D$ and
$D_s$ dependence of all integral coefficients in the sunrise hierarchy in
\fig{fig:diagSunrise}. All these results were used in the first
calculation of a two-loop amplitude in the framework of
numerical unitarity~\cite{4g2loop}.

A systematic study of the efficiency and stability of the numerical unitarity
approach is left to future work, however, we have observed that the algorithm
appears sufficiently fast and numerically stable. We hope in the future to
explore the use of the numerical unitarity method to two-loop amplitudes with
more than four external particles.

The algorithm we presented to deal with subleading-pole contributions requires to
handle an
enlarged set of linear equations when computing integral coefficients.
It would be interesting to explore alternative approaches which organize
the cut equations more effectively.  For example, it is possible to directly
associate subleading contributions to subtracted gluon amplitudes to tree
amplitudes with graviton exchange~\cite{Stieberger:2015kia}. Further ideas
include accessing the subleading poles through numerical limits, or using so-called BCJ
relations~\cite{Bern:2008qj} to relate the subleading contributions to other
color-ordered cuts.

\section*{Acknowledgments}

We thank Z.~Bern, A. de Freitas and D.A.~Kosower for helpful discussions. We
particularly thank Z.~Bern for providing analytical expressions
from ref.~\cite{fourgluons}.
S.A.'s work is supported by the Juniorprofessor Program of Ministry of Science, 
Research and the Arts of the state of Baden-W\"urttemberg, Germany.
H.I.'s work is supported by a Marie Sk{\l}odowska-Curie
Action Career-Integration Grant PCIG12-GA-2012-334228 of the European Union.
The work of F.F.C., M.J.
and B.P. is supported by the Alexander von Humboldt Foundation, in the framework
of the Sofja Kovalevskaja Award 2014, endowed by the German Federal Ministry of
Education and Research.  This work was performed on the bwUniCluster funded by
the Ministry of Science, Research and the Arts Baden-W\"urttemberg and the
Universities of the State of Baden-W\"urttemberg, Germany, within the framework
program bwHP.  The authors are grateful to the Mainz Institute for Theoretical
Physics (MITP) for its hospitality and its partial support during the
completion of this work.


\begin{thebibliography}{99}

\bibitem{Unitarity}
Z.~Bern, L.~J.~Dixon, D.~C.~Dunbar and D.~A.~Kosower,
``One-loop $n$-point gauge theory amplitudes, unitarity
 and collinear limits,''
Nucl.\ Phys.\ B {\bf 425}, 217 (1994)
[hep-ph/9403226];
%
``Fusing gauge theory tree amplitudes into loop amplitudes,''
Nucl.\ Phys.\ B {\bf 435}, 59 (1995)
[hep-ph/9409265];
%
Z.~Bern, L.~J.~Dixon and D.~A.~Kosower,
``One-loop amplitudes for e+ e- to four partons,''
Nucl.\ Phys.\  B {\bf 513}, 3 (1998)
[hep-ph/9708239];
R.~Britto, F.~Cachazo and B.~Feng,
``Generalized unitarity and one-loop amplitudes in N = 4  super-Yang-Mills,''
Nucl.\ Phys.\  B {\bf 725}, 275 (2005)
[hep-th/0412103].

\bibitem{OPP} 
  G.~Ossola, C.~G.~Papadopoulos and R.~Pittau,
  ``Reducing full one-loop amplitudes to scalar integrals at the integrand level,''
  Nucl.\ Phys.\ B {\bf 763}, 147 (2007)
  [hep-ph/0609007].

\bibitem{NumUnitarity}
 R.~K.~Ellis, W.~T.~Giele and Z.~Kunszt,
 ``A Numerical Unitarity Formalism for Evaluating One-Loop Amplitudes,''
 JHEP {\bf 0803} (2008) 003
 [arXiv:0708.2398 [hep-ph]].

\bibitem{GKM}
W.~T.~Giele, Z.~Kunszt and K.~Melnikov,
``Full one-loop amplitudes from tree amplitudes,''
JHEP {\bf 0804}, 049 (2008)
[arXiv:0801.2237 [hep-ph]].

\bibitem{BlackHat}
C.~F.~Berger, Z.~Bern, L.~J.~Dixon, F.~Febres~Cordero, D.~Forde, H.~Ita,
D.~A.~Kosower and D.~Ma\^{\i}tre,
``An Automated Implementation of On-Shell Methods for One-Loop
Amplitudes,''
Phys.\ Rev.\ D {\bf 78}, 036003 (2008)
[arXiv:0803.4180 [hep-ph]].

\bibitem{DPropAnalytic}
J.~H.~Zhang,
 ``Multidimensional Residues for Feynman Integrals with Generic Power of Propagators,''
 arXiv:1112.4136 [hep-th];
  P.~Mastrolia, E.~Mirabella, G.~Ossola and T.~Peraro,
  ``Multiloop Integrand Reduction for Dimensionally Regulated Amplitudes,''
  Phys.\ Lett.\ B {\bf 727}, 532 (2013)
  [arXiv:1307.5832 [hep-ph]];
M.~Sogaard and Y.~Zhang,
 ``Unitarity Cuts of Integrals with Doubled Propagators,''
 JHEP {\bf 1407} (2014) 112
 [arXiv:1403.2463 [hep-th]].

\bibitem{fourgluons} 
  Z.~Bern, A.~De Freitas and L.~J.~Dixon,
  ``Two loop helicity amplitudes for gluon-gluon scattering in QCD and supersymmetric Yang-Mills theory,''
  JHEP {\bf 0203}, 018 (2002)
  [hep-ph/0201161];

\bibitem{UnitarityAnalytics2loop} 
  Z.~Bern, L.~J.~Dixon and D.~A.~Kosower,
  ``A Two loop four gluon helicity amplitude in QCD,''
  JHEP {\bf 0001} (2000) 027
  [hep-ph/0001001].
%
  S.~Badger, H.~Frellesvig and Y.~Zhang,
  ``A Two-Loop Five-Gluon Helicity Amplitude in QCD,''
  JHEP {\bf 1312}, 045 (2013)
  [arXiv:1310.1051 [hep-ph]];
  S.~Badger, G.~Mogull, A.~Ochirov and D.~O'Connell,
  ``A Complete Two-Loop, Five-Gluon Helicity Amplitude in Yang-Mills Theory,''
  JHEP {\bf 1510}, 064 (2015)
 [arXiv:1507.08797 [hep-ph]];
  D.~C.~Dunbar and W.~B.~Perkins,
  ``Two-loop five-point all plus helicity Yang-Mills amplitude,''
  Phys.\ Rev.\ D {\bf 93}, no. 8, 085029 (2016)
  [arXiv:1603.07514 [hep-th]];
  D.~C.~Dunbar, G.~R.~Jehu and W.~B.~Perkins,
  ``The two-loop n-point all-plus helicity amplitude,''
  Phys.\ Rev.\ D {\bf 93}, no. 12, 125006 (2016)
  [arXiv:1604.06631 [hep-th]].

\bibitem{FDH} 
  Z.~Bern, A.~De Freitas, L.~J.~Dixon and H.~L.~Wong,
  ``Supersymmetric regularization, two loop QCD amplitudes and coupling shifts,''
  Phys.\ Rev.\ D {\bf 66}, 085002 (2002)
  [hep-ph/0202271].

\bibitem{NumUnitarity2}
  H.~Ita,
  ``Two-loop Integrand Decomposition into Master Integrals and Surface Terms,''
  Phys.\ Rev.\ D {\bf 94} (2016) no.11,  116015
[arXiv:1510.05626 [hep-th]].

\bibitem{Badger:2016ozq} 
  S.~Badger, G.~Mogull and T.~Peraro,
  ``Local integrands for two-loop all-plus Yang-Mills amplitudes,''
  JHEP {\bf 1608}, 063 (2016)
  [arXiv:1606.02244 [hep-ph]].

\bibitem{IBPGKK}
J.~Gluza, K.~Kajda and D.~A.~Kosower,
``Towards a Basis for Planar Two-Loop Integrals,''
Phys.\ Rev.\ D {\bf 83} (2011) 045012
[arXiv:1009.0472 [hep-th]].

\bibitem{BGrec}
F.~A.~Berends and W.~T.~Giele,
``Recursive Calculations for Processes with n Gluons,''
Nucl.\ Phys.\ B {\bf 306} (1988) 759.

\bibitem{Ochirov:2016ewn}
  A.~Ochirov and B.~Page,
  ``Full Colour for Loop Amplitudes in Yang-Mills Theory,''
  JHEP {\bf 1702} (2017) 100
  [arXiv:1612.04366 [hep-ph]].

\bibitem{Smirnov:2014hma} 
  A.~V.~Smirnov,
  ``FIRE5: a C++ implementation of Feynman Integral REduction,''
  Comput.\ Phys.\ Commun.\  {\bf 189}, 182 (2015)
  [arXiv:1408.2372 [hep-ph]].

\bibitem{tensorBasis}
  P.~Mastrolia and G.~Ossola,
  ``On the Integrand-Reduction Method for Two-Loop Scattering Amplitudes,''
  JHEP {\bf 1111} (2011) 014
  [arXiv:1107.6041 [hep-ph]]; 
  S.~Badger, H.~Frellesvig and Y.~Zhang,
  ``Hepta-Cuts of Two-Loop Scattering Amplitudes,''
  JHEP {\bf 1204} (2012) 055
  [arXiv:1202.2019 [hep-ph]]; 
  Y.~Zhang,
  ``Integrand-Level Reduction of Loop Amplitudes by Computational Algebraic Geometry Methods,''
  JHEP {\bf 1209} (2012) 042
  [arXiv:1205.5707 [hep-ph]]; 
  P.~Mastrolia, E.~Mirabella, G.~Ossola and T.~Peraro,
  ``Scattering Amplitudes from Multivariate Polynomial Division,''
  Phys.\ Lett.\ B {\bf 718} (2012) 173
  [arXiv:1205.7087 [hep-ph]].

\bibitem{4g2loop} 
  S.~Abreu, F.~Febres Cordero, H.~Ita, M.~Jaquier, B.~Page and M.~Zeng,
  ``Two-Loop Four-Gluon Amplitudes with the Numerical Unitarity Method,''
  arXiv:1703.05273 [hep-ph].


\bibitem{Stieberger:2015kia} 
  S.~Stieberger and T.~R.~Taylor,
  ``Subleading terms in the collinear limit of Yang-Mills amplitudes,''
  Phys.\ Lett.\ B {\bf 750}, 587 (2015)
  [arXiv:1508.01116 [hep-th]].

\bibitem{Bern:2008qj} 
  Z.~Bern, J.~J.~M.~Carrasco and H.~Johansson,
  ``New Relations for Gauge-Theory Amplitudes,''
  Phys.\ Rev.\ D {\bf 78}, 085011 (2008)
 [arXiv:0805.3993 [hep-ph]].


\end{thebibliography}
\end{document}